\documentclass[a4paper]{article}

\usepackage{INTERSPEECH2022}
\usepackage{comment}
\usepackage{multirow}
\usepackage{color}
\usepackage[ruled]{algorithm}
\usepackage{algpseudocode}
\usepackage{caption} 
\usepackage{cite} 
\usepackage{subcaption}
\usepackage{amssymb}
\usepackage{enumitem}
\usepackage{color}
\usepackage{bibspacing}
\setlist[itemize]{align=parleft,left=0pt..1em}
\title{Conformer Based Elderly Speech Recognition System\\ for Alzheimer's Disease Detection}
\name{Tianzi Wang, Jiajun Deng, Mengzhe Geng, Zi Ye, Shoukang Hu, Yi Wang, Mingyu Cui, Zengrui Jin, Xunying Liu, Helen Meng}
\address{
  The Chinese University of Hong Kong, Hong Kong SAR, China}
\email{\footnotesize{\{twang,jjdeng,mzgeng,zye,skhu,ywang,mycui,zrjin,xyliu,hmmeng\}@se.cuhk.edu.hk}}

\begin{document}
\bstctlcite{IEEEexample:BSTcontrol}

\maketitle
\begin{abstract}

Early diagnosis of Alzheimer’s disease (AD) is crucial in facilitating preventive care to delay further progression. This paper presents the development of a state-of-the-art Conformer based speech recognition system built on the DementiaBank Pitt corpus for automatic AD detection. The baseline Conformer system trained with speed perturbation and SpecAugment based data augmentation is significantly improved by incorporating a set of purposefully designed modeling features, including neural architecture search based auto-configuration of domain-specific Conformer hyper-parameters in addition to parameter fine-tuning; fine-grained elderly speaker adaptation using learning hidden unit contributions (LHUC); and two-pass cross-system rescoring based combination with hybrid TDNN systems. An overall word error rate (WER) reduction of 13.6\% absolute (34.8\% relative) was obtained on the evaluation data of 48 elderly speakers. Using the final systems’ recognition outputs to extract textual features, the best-published speech recognition based AD detection accuracy of 91.7\% was obtained. 

\noindent\textbf{Index Terms}: Automatic Speech Recognition, Elderly Speech,
Alzheimer's Disease Detection, Conformer, Dementia

\end{abstract}

\section{Introduction}
\label{sec:introduction}
Despite the rapid progress of automatic speech recognition (ASR) technologies in the past few decades, accurate recognition of elderly and disordered speech remains a challenging task \cite{christensen2013combining,sehgal2015model,yu2018development,xiong2020source,liu2020exploiting,ye2021development,geng2022spectro,deng2021bayesian,geng2020invest}. Ageing presents enormous challenges to health care worldwide. Neurocognitive disorders, such as Alzheimer’s disease (AD), are often found among older adults \cite{alzheimer20192019} and manifest themselves in speech and language impairments \cite{fraser2016linguistic,konig2018fully}. ASR-based assistive technology development tendering for such users’ needs plays a vital role in not only improving their quality of life and social inclusion, but also facilitating large scale automatic speech-based early diagnosis of neurocognitive impairment and preventive care \cite{ferri2005global}. As a non-intrusive, automatic, more scalable, and less costly alternative to other screening techniques based on brain scans or blood tests, there has been increasing interest in developing speech-based AD diagnosis systems, in particular during the recent ADReSS challenge \cite{luz2020alzheimer,luz2021detecting}. For these systems, linguistic features extracted from the elderly speech transcripts play a key role \cite{vipperla2010ageing,rudzicz2014speech,zhou2016speech,konig2018fully,mirheidari2019dementia,toth2018speech,ye2021development,li2021comparative,pan21c_interspeech,yuan2020disfluencies,syed2021automated,cummins2020comparison,searle2020comparing,zhu2021wavbert,gauder2021alzheimer,balagopalan2021comparing,abulimiti2020automatic}. To this end, accurate recognition of elderly speech recorded during neurocognitive impairment assessment interviews is crucial.

Elderly speech brings challenges on all fronts to current deep learning based ASR technologies predominantly targeting non-aged, healthy adult users. First, a large mismatch between such data and non-aged adult voices is often observed. Such difference manifests itself across many fronts including articulatory imprecision, decreased volume and clarity, changes in pitch, increased dysfluencies and slower speaking rate \cite{hixon1964restricted, kent2000dysarthrias}. Second, the co-occurring disabilities, mobility, or accessibility limitations often found among elderly speakers lead to the difficulty in collecting large quantities of such data that are essential for current data intensive ASR system development. In addition, sources of variability commonly found in normal speech including accent or gender, when further compounded with those over age and speech and language pathology severity, create large diversity among elderly speakers \cite{kodrasi2020spectro,smith1987temporal}. 

Inspired by the successful application of Convolution-augmented Transformer (Conformer) end-to-end models to a wide range of normal speech recognition task domains \cite{gulati2020conformer,guo2021recent}, this paper presents the development of the first Conformer based elderly speech recognition system on the largest publicly available English DementiaBank Pitt corpus \cite{becker1994natural} for AD detection. In order to address the above challenges in elderly speech recognition, the baseline Conformer system trained with speed perturbation and SpecAugment based data augmentation is significantly improved by incorporating a set of purposefully designed modeling features. First, in addition to conventional cross-domain parameter fine-tuning of normal speech pre-trained systems, their large mismatch against elderly speech is addressed using neural architecture search (NAS) based auto-configuration of domain-specific Conformer hyper-parameters, for example, the convolution kernel size used to encode the acoustic temporal context span in elderly speech utterances characterized by increased dysfluencies and shorter length, akin to the context offsets previously observed on hybrid TDNN systems \cite{hu2022neural,deng2021bayesian}. Second, fine-grained elderly user personalization of Conformer models is performed by learning hidden unit contributions (LHUC) \cite{swietojanski2016learning,deng2021confider}. Lastly, the cross-system complementarity between hybrid TDNN and end-to-end Conformer based ASR systems is further exploited using a two-pass cross-system rescoring approach \cite{watanabe2017hybrid,sainath2019two,li2019integrating,myc2021twopass}.

An overall word error rate (WER) reduction of 13.6\% absolute (34.8\% relative) was obtained on the DementiaBank Pitt evaluation data (subsuming the ADReSS test data based on the same speakers but only the picture description task) consisting of 48 elderly speakers. Using the final systems’ recognition outputs to extract textual features, the best-published speech recognition based AD detection accuracy of 91.7\% was obtained. An analysis of the correlation between speech recognition accuracy and AD detection performance is further presented. 

The main contributions of this paper are summarized below. To the best of our knowledge, this is the first work to design state-of-the-art Conformer based ASR systems tailored for elderly speech recognition and downstream AD diagnosis tasks. 
In contrast, the previous research used either off-shelf commercial speech recognition systems \cite{konig2018fully, pompili2020pragmatic,fraser2013automatic,zhu2021wavbert,gauder2021alzheimer,balagopalan2021comparing}, or more traditional GMM-HMM or hybrid DNN models in system development \cite{mirheidari2019dementia,rudzicz2014speech,lehr2012fully,zhou2016speech,ye2021development,abulimiti2020automatic}. 
In addition, this paper presents the first use of multi-pass rescoring based system combination approaches for hybrid TDNN and end-to-end Conformer based ASR systems. In contrast, prior researches on hybrid and end-to-end system combination were mainly conducted in the context of non-Conformer based architectures such as CTC, LAS, and RNN transducers \cite{watanabe2017hybrid,sainath2019two,li2019integrating}.

The rest of this paper is organized as follows. Section 2 introduces the data and the baseline Conformer ASR system used. Section 3 presents NAS based auto-configuration of Conformer hyper-parameters, domain and speaker adaptation and system combination approaches. Section 4 shows the AD detection system performance using ASR outputs. Finally, the conclusions are drawn and future works are discussed in Section 5. 

\begin{figure*}[htb]
    \centering
    \includegraphics[width=\textwidth]{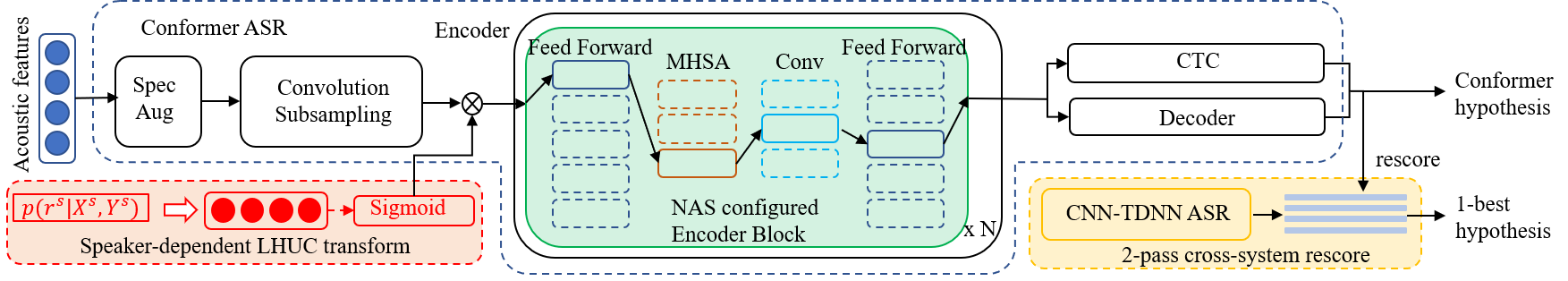}
    \caption{\rm The overall Conformer based elderly speech recognition system architecture with neural architectural search auto-configured encoder block hyper-parameters (green, centre), LHUC speaker adaptation (red, left bottom) and 2-pass decoding based system combination with a hybrid CNN-TDNN system (yellow, right bottom) producing the final ASR system outputs. }
    \label{fig:my_label}
\end{figure*}

\section{Task Description}
This section describes the audio and text data used in this paper and the baseline ASR system.

\noindent\textbf{Audio Data}: The English DementiaBank Pitt corpus \cite{becker1994natural} contains 33 hours of speech audio recorded over interviews between the 292 elderly participants and the clinical investigators. It is further split into a 27.2-hour training set, a 4.8-hour development set and a 1.1-hour evaluation set for ASR system development. After silence stripping \cite{ye2021development}, the training set contains 15.7 hours of audio data (29682 utterances) while the development and evaluation sets contain 2.5 hours (5103 utterances) and 0.6 hours (928 utterances) of audio respectively. After a combination of speaker independent of elderly speech and dependent speed perturbation \cite{ye2021development} of non-aged investigators’ speech based data augmentation, the duration of training data was increased to 58.9 hours (112830 utterances). The evaluation set is exactly based on the same 48 speakers’ Cookie section recordings as the ADReSS\cite{luz2020alzheimer} test set, while the development set contains the remaining recordings of the same speakers in other task sections if available. 

\noindent\textbf{Text Data}: For language models, the transcripts (167k words) of the Pitt data \cite{becker1994natural} was used to construct word level 4-gram language models (LMs) with modified Kneser-Ney smoothing using the SRILM toolkit \cite{stolcke2002srilm} to rescore Conformer N-best recognition (N=100) outputs. A 3.6k word recognition vocabulary covering all the words in the Pitt corpus was used. 

\noindent\textbf{Baseline System}: Conformer sequence trained end-to-end ASR models were built. Following the ESPnet \cite{watanabe2018espnet} recipe setup, the encoder contains 2 Convolution blocks to downsample the 40-dimension Mel-scale filter banks (FBKs) inputs, followed by 12 Conformer blocks. Inside each block, the feedforward layer dimensionality, the number and dimensions of attention heads were set to 2048, 4 and 256 respectively. Relative position embedding was also used. The convolution kernel size was set as 31. The decoder contains 6 Transformer blocks with the number and dimensions of attention heads were set to 4 and 256 respectively. The output vocabulary includes 26 English characters, a space token, an apostrophe and a special CTC blank symbol. An interpolated CTC+AED (weighting 3:7) cost function was used in model training on NVIDIA V100 GPUs. Matched pairs sentence-segment word error (MAPSSWE) based statistical significance test was performed at a significance level $\alpha=0.05$. 
\vspace{-0.4cm}
\section{Conformer Based Elderly Speech Recognition System}
This section presents the performance of the baseline Conformer systems before introducing a series of techniques to further improve the recognition accuracy. The overall architecture is shown in Figure 1. 

\subsection{Baseline and Manually designed System Performance}

The performance of the baseline Conformer system configured using the ESPnet recipe\footnote{ESPnet: egs/swbd/asr1/run.sh} and trained on the data augmented 59 hour Pitt corpus is shown in the first line (Sys. 1) in Table 1. Motivated by the domain specific sensitivity over TDNN hyper-parameters previously studied in \cite{deng2021bayesian}, a series of manual re-configuration of Conformer hyper-parameters were performed to improve its performance. By ablation studies over varying the number of encoder and decoder Transformer layers and the resulting impact on performance, the optimized number of decoder Transformer blocks was increased from 6 to 12. 
Similarly, the convolution kernel size was manually tuned and reduced from 31 to 7. The resulting manually designed Conformer system outperformed the baseline Switchboard recipe configured system (Sys. 2 vs. Sys. 1, Table 1) by statistically significant a WER reduction of 1.4\% absolute (1.5\% absolute for participants in the evaluation set) on average, while the number of model parameters increased from 42.3M to 51.8M. 

\begin{table*}[tb]\caption{\label{tab:table-darts}\textnormal{Performance (WER\%, \#Params) of baseline, manually and NAS auto-configured Conformer models derived using Softamx or Gumbel-Softmax (Gumbel) DARTS of Section 3.2. The manually designed Conformer (Sys. 2) serves as the start point of a progressive architecture search performed in turn over three types of hyper-parameters of Conformer within each encoder block: a) the dimensionality of feedforward and macron-feedforward layers "FD" (in bracketed pair), where the dimensionality indices denote a choice from \{512, 1024, 2048, 3072, 4096\}; b) the number of attention heads "AH"; and c) the kernel size of the convolution module "CK". Their respective search scope and 1-best hyper-parameter configurations (separated by ";" between layers) in Col. 3 and 4. The IDs of systems auto-configured with best hyper-parameters learned at each of three stages of NAS marked in bold with "$\ast$". $\eta$ is the penalty factor penalized DARTS of Eqn.3. $\dag$ denotes a statistically significant difference obtained over the baseline system (Sys. 1). }}

\scriptsize
\renewcommand\arraystretch{1.3}
\begin{center}
\vspace{-0.4cm}
\scalebox{0.9}[0.9]{
\begin{tabular}{c|c|l|c|c|c|cc|cc|c|l}
\hline
Sys. & NAS                     & \begin{tabular}[c]{@{}l@{}}Hyper- \\Param.\end{tabular}                       & \begin{tabular}[c]{@{}l@{}}Search \\ Scope\end{tabular}                                     & NAS configured Encoder Hyper-parameters& $\eta$             & \multicolumn{2}{c|}{Dev}                           & \multicolumn{2}{c|}{Eval}                          & All & \#Params \\ \cline{7-10}
    &                         &                                                                                          &                                                                                             &                                                        &                    & \multicolumn{1}{c}{Inv} & \multicolumn{1}{c|}{Par} & \multicolumn{1}{c}{Inv} & \multicolumn{1}{c|}{Par} &         &          \\ \hline \hline
1   & Baseline               & \multirow{2}{*}{-}      & \multirow{2}{*}{-}       &  FD: (2:2)$\times$12; AH: (4)$\times$12; CK:(15)$\times$ 12                                     & \multirow{2}{*}{-} &  21.9                       & 50.2                         &  18.3                       &  39.1                        &  35.9       &     42.3M     \\ \cline{1-1} 
2   & Manual               &      &            &   FD: (2:2)$\times$12; AH: (4)$\times$12; CK: (7)$\times$12  &                    &                         21.0$\dag$&                        48.2$\dag$ &                       19.4  &          37.6  &        34.5$\dag$ &         51.8M \\ \hline \hline
3   & Softmax                 & \multirow{3}{*}{\begin{tabular}[c]{@{}l@{}}FD\end{tabular}} & \multirow{3}{*}{\begin{tabular}[c]{@{}l@{}}\{0, 1, 2, \\3, 4\}\end{tabular}} & \begin{tabular}[c]{@{}c@{}}(1,2);(0,4);(1,1);(4,0);(1,4);(2,2);(4,4);(2,3);(2,3);(1,4);(0,1);(1,4) \end{tabular}  & \multirow{2}{*}{0} &                         21.3$\dag$&                         \textbf{47.5$\dag$} &                         19.0&           37.8&         34.4$\dag$&         54.1M \\ \cline{1-2} \cline{5-5}
4   & \multirow{2}{*}{Gumbel} &   &      &  \begin{tabular}[c]{@{}c@{}}(0,4);(2,4);(4,3);(4,1);(0,4);(2,4); (1,1);(3,0);(2,3);(1,1);(0,1);(1,4)\end{tabular}    &       &                       21.4$\dag$  &                      48.6$\dag$    &                      19.8 &                       38.0&         34.9$\dag$&          53.9M\\ \cline{1-1} \cline{5-6} 
\textbf{5*}   &                         &           &    &  \begin{tabular}[c]{@{}c@{}} (1,1);(1,3);(0,3);(0,2);(0,0);(0,0);(0,0);(0,0);(0,0);(0,0);(0,1);(1,0)\end{tabular}    & 0.03               &                        \textbf{20.8$\dag$} &                      48.2$\dag$ &                      \textbf{19.0}&                       \textbf{37.6$\dag$}&         \textbf{34.4$\dag$}&          37.6M\\     \hline \hline
6  & Softmax                 & \multirow{3}{*}{\begin{tabular}[c]{@{}l@{}}+AH\end{tabular}}  &   \multirow{3}{*}{\{2, 4, 8\}}  & 8;8;8;8;8;8;4;2;2;8;8;8  & \multirow{2}{*}{0} &                     21.3$\dag$  &                        48.7$\dag$ &                      \textbf{16.5}  &                        37.7$\dag$ &        34.7$\dag$ & 40.2M        \\ \cline{1-2}  \cline{5-5}
7  & \multirow{2}{*}{Gumbel} &    &              &  8;8;8;8;8;8;4;2;2;8;8;8  &                    &                        21.3$\dag$ &                        48.7$\dag$  &                      \textbf{16.5}  &                        37.7$\dag$ &       34.7$\dag$ &  40.2M       \\ \cline{1-1} \cline{5-6}
\textbf{8*}  &                         &       &       &   8;8;4;4;8;8;8;2;2;4;8;8   & 0.03               &                       \textbf{20.9$\dag$} &                       \textbf{47.9$\dag$} &                       17.6  &                        \textbf{37.6$\dag$}  &       \textbf{34.3$\dag$} &  39.5M        \\  \hline \hline
9  & Softmax                 & \multirow{4}{*}{\begin{tabular}[c]{@{}l@{}}\,\,+CK\end{tabular}}   & \multirow{4}{*}{\{3,5,7\}}                                                                    & 7;7;7;7;5;7;7;7;7;7;7;7     & \multirow{2}{*}{0} &  21.2$\dag$                        &  \textbf{47.7$\dag$}                        & \textbf{16.4$\dag$}                        &  38.1                        &  34.4$\dag$       &  39.5M        \\ \cline{1-2}  \cline{5-5}
10  & \multirow{3}{*}{Gumbel} &                                                                                         &                                                                                             &       7;7;3;7;5;7;5;7;7;7;5;7           &                    &  21.3$\dag$                       &   48.4$\dag$                       & 19.3                        & 37.2$\dag$                         &  34.6$\dag$       &  39.5M        \\ \cline{1-1} \cline{5-6}
11  &                         &     &                                   &    7;7;3;7;5;7;5;7;7;7;5;7      & 0.03               & 21.3$\dag$                       & 48.4$\dag$                         & 19.3                       &  37.2$\dag$                        &     34.6$\dag$   &  39.5M   \\ \cline{1-1}\cline{5-6}
\textbf{12*}  &                         &                                                                                          &                                                                                             &   7;7;3;7;5;5;7;7;7;7;7;7 & 0.3               & \textbf{21.0$\dag$}                       & 47.9$\dag$                         & 19.0                       &  \textbf{36.7$\dag$}                        &     \textbf{34.2$\dag$}   &  39.5M   \\\cline{1-12}
\end{tabular}
}
\end{center}
\vspace{-0.9cm}
\end{table*}
\vspace{-0.1cm}
\subsection{Neural Architecture Search}
Designing and evaluating suitable neural network architectures for specific task domains requires a large effort of human experts and is extremely expensive. To automatically learn the suitable Conformer architectural design for the target elderly speech domain, differentiable neural architecture search (DARTS) \cite{liu2018darts} was used to further optimize three groups hyper-parameters inside each of the 12 Conformer encoder blocks: a) the feedforward layer dimensionality; b) the number of attention heads; and c) the convolution kernel size. These are highlighted in Figure 1 (green box, centre). The manually designed Conformer (Sys. 2, Table 1) serves as the start point of NAS.
The general form of DARTS architecture selection is as follows: 
\vspace{-0.1cm}
\begin{equation}
\vspace{-0.1cm}
     x^l=\sum_{i=1}^{N^l}\lambda_{i}^{l}c_i^l(x^{l-1})=\sum_{i=1}^{N^l}\frac{\text{exp}(\alpha_i^{l})}{\sum_{j=1}^{N^l}\text{exp}(\alpha_j^{l})}c_i^l(x^{l-1})
\end{equation}
where $c_i^l$ and $\lambda_i^l$ are the i-th candidate architecture choice of l-th layer and its corresponding weight respectively. $\lambda_i^l$ is modelled by a Softmax function over a vector $\alpha^l$, whose dimensionality equals to the total number of candidate architectures, $N^{l}$. The estimation of standard network parameters excluding the architectural parameters $\alpha_i^{l}$ inside the 
super-network model is decoupled from that of the architecture parameters $\alpha_i^{l}$ \cite{hu2022neural, guo2020single}. This leads to the pipelined DARTS allowing the architecturel weights to be learned on separate held-out data. The optimal architecture with the largest weight is selected. 

\noindent\textbf{Gumbel-Softmax DARTS}:
For traditional DARTS methods, when similar architecture weights are obtained using a flattened Softmax function, the confusion over different candidate systems increases and search errors may occur. To this end, a Gumbel-Softmax distribution \cite{xie2018snas, hu2022neural} is used to sharpen the architecture weights to produce approximately a one-hot vector. This allows the confusion between different architectures to be minimised. The architecture weights are computed as,
\vspace{-0.1cm}
\begin{equation}
\vspace{-0.1cm}
     \lambda_{i}^{l}=\frac{\text{exp}(\text{log}(\alpha_i^{l}+G_{i}^l)/T)}{\sum_{j=1}^{N^l}\text{exp}(\text{log}(\alpha_j^{l}+G_j^{l})/T)}
\end{equation}
where $G_{i}^{l}=\text{-log(-log}(U_i^l))$ is the Gumbel variables and and $U_i^l$ is a uniform random variable. As the temperature parameter $T$ decreases to zero, Eqn.2  approaches a categorical distribution.

\noindent\textbf{Penalized DARTS}:
In order to avoid over-parameterized during architecture search, a penalty loss incorporating the number of parameters for each candidate choice was jointly optimized with the original Conformer training loss function: 

\begin{equation}
     \mathcal{L}=\mathcal{L}_{Conformer} + \eta\sum\nolimits_{i, l}\alpha_{i}^{l}P_{i}^{l}
\end{equation}

where $P_{i}^{l}$ is the number of parameters of the i-th candidate architecture at the l-th layer, and $\eta$ is the penalty scaling factor  empirically adjusted for performance vs. complexity trade-off. 

Several trends can be found in the results of Table 1. First, the manually configured Conformer system (Sys. 2) outperformed the baseline ESPnet recipe Conformer (Sys. 1) by statistically significant WER reductions of 1.4\% absolute on average across the two test sets. However, the doubling of the number of decoder layers from 6 to 12 in the manually configured system (Sys. 2) led to a 22.5\% increase in parameters. Second, the use of penalized Gumbel-Softmax DARTS consistently produced the most compact and best performing system architecture at each stage of NAS search (Sys. 5 vs. Sys. 3 \& 4, Sys. 8 vs. Sys. 6 \& 7, and Sys. 12 vs. Sys. 9-11). The largest architectural compression of 27.4\% was obtained on optimizing the feedforward layers' dimensionality (Sys. 5 vs. 2) while incurring no performance degradation. Third, without applying model size penalty ($\eta=0$), the Softmax and Gumbel-Softmax DARTS configured systems produced comparable performance and system complexity. Lastly, the best performing penalized Gumbel-Softmax DARTS auto-configured system (Sys. 12) produced statistically significant WER reductions of 1.7\% absolute (2.4\% absolute for participants’ data in the evaluation set) with 6.6\% fewer parameters compared to the baseline system (Sys.1).
\vspace{-0.1cm}
\subsection{Domain Adaptation}
In order to exploit large quantities of out-of-domain, non-aged adult speech pre-trained Conformer systems, cross-domain adaptation was considered. A 960-hour LibriSpeech corpus trained Conformer models that were either manually designed or NAS auto-configured (comparable to Sys. 2 \& 12 in Table 1) were cross-domain adapted to the 59-hour in-domain Pitt data after speed perturbation. During domain adaptation, the projection layers of the Conformer CTC and decoder modules were removed and replaced with a randomly initialized Softmax layer, while the rest of the system initialized using the LibriSpeech pre-trained model, before fine-tuning to convergence\footnote{Ablation studies suggest alternative cross-domain adaptation settings involving other combinations among: a) parameter random re-initialization followed by re-estimation; b) fixing the pre-trained parameters; or c) fine-tuning the pre-trained parameters all led to performance degradation.}.

As shown in Table 2, irrespective of the architecture design being manual or NAS configuration, cross-domain adaptation of normal, non-aged speech pre-trained Conformer systems consistently produced statistically significant WER reductions of 8.0\%-8.9\% absolute (23.4\%-25.8\% relative) on average across both test sets (Sys. 13, 15 vs. Sys. 2, 12). 
It is also worth noting that the precise performance ranking between the manually designed or NAS auto-configured systems has changed after domain adaptation. This may be due to the fact that NAS configured hyper-parameters were learned using the DementiaBank Pitt data only, while being sub-optimal when used to configure the out-of-domain LibriSpeech Conformer pre-training. Similar domain sensitivity over TDNN hyper-parameters was previously studied in \cite{deng2021bayesian} and will be investigated in future research.

\vspace{-0.2cm}
\subsection{Speaker Adaptation}
Individuals experiencing AD at different stages of progression exhibit highly diverse voice characteristics. To this end, speaker adaptation techniques play a central role in reducing the mismatch between ASR systems and target elderly users. In order to model the large variability among elderly participants in the Pitt data, fine-grained elderly user personalization of Conformer models is further performed by learning hidden unit contributions (LHUC) \cite{swietojanski2016learning, deng2021confider}. As Table 2 shows, the speaker adapted Conformer systems (Sys. 14, 16) outperformed the comparable domain adapted, but speaker independent Conformer systems (Sys.13, 15) by 0.3\% absolute (0.3\% absolute for participants in the evaluation set) in WER reduction. 

\begin{table}[tb]
\caption{\label{tab:table-adpt}\textnormal{Performance of domain and speaker adapted the manually designed and NAS auto-configured Conformer systems (on top of Sys. 1, 2 \& 12 in Table 1). † denotes a statistically significant difference obtained over the baseline system (Sys. 1). }}
\begin{center}
\vspace{-0.5cm} 
\scalebox{0.7}[0.7]{
\begin{tabular}{c|c|p{5mm}p{5mm}|p{6mm}p{6mm}p{6mm}p{6mm}|l}
\hline
\multirow{2}{*}{Sys.} & \multirow{2}{*}{NAS}   & \multicolumn{2}{c|}{Adaption}                      & \multicolumn{2}{c|}{Dev}                           & \multicolumn{2}{c|}{Eval}                          & \multirow{2}{*}{All} \\ \cline{3-8}
                      &  & \multicolumn{1}{p{3mm}}{Dom} & \multicolumn{1}{p{3mm}|}{Spk} & \multicolumn{1}{c}{Inv} & \multicolumn{1}{c|}{Par} & \multicolumn{1}{c}{Inv} & \multicolumn{1}{c|}{Par} &                      \\ \hline \hline
1 & Baseline     & $\times$                       & $\times$                        &21.9                         & 50.2                         & 18.3                        & 39.1                         & 35.9             \\ \hline
2 & \multirow{3}{*}{Manual}     & $\times$                       & $\times$                        &21.0$\dag$                         & 48.2$\dag$                         & 19.4                        & 37.6                         & 34.5$\dag$                     \\ 
13 &                      & $\checkmark$            & $\times$                        &       \textbf{16.0}$\dag$            &             35.8$\dag$         & \textbf{15.2}$\dag$                    & 26.7$\dag$                     &           25.6$\dag$          \\
 14 &                      & $\checkmark$            & $\checkmark$             &          \textbf{16.0}$\dag$            &           \textbf{35.2}$\dag$            &        15.3$\dag$            &                \textbf{26.4}$\dag$          &       \textbf{25.3}$\dag$             \\ \hline
 12 & \multirow{3}{*}{Gumbel} & $\times$                       & $\times$                        & 21.0$\dag$                        &  47.9$\dag$                        &  19.0                       & 36.7$\dag$                         &  34.2$\dag$                    \\
 15 &                      & $\checkmark$            & $\times$                        &          \textbf{16.1}$\dag$               &          36.5$\dag$                &          16.0$\dag$   &            27.9$\dag$              &            26.2$\dag$            \\
 16 &                      & $\checkmark$            & $\checkmark$             &       \textbf{16.1}$\dag$                &        \textbf{36.1}$\dag$                &    \textbf{14.7}$\dag$               &         \textbf{27.6}$\dag$              &        \textbf{25.9}$\dag$          \\
\hline
\end{tabular}
}
\vspace{-0.6cm} 
\end{center}
\end{table}

\begin{table}[]
\caption{\label{tab:table-comb}\textnormal{Performance of system combination between end-to-end Conformer and hybrid CNN-TDNN models derived using two-pass decoding. "CNN-TDNN $\rightarrow$ Y" denotes CNN-TDNN system produced N-best outputs in a $1^{\text{st}}$ decoding pass prior to $2^{\text{ed}}$ pass rescoring by Sys. Y using cross-system score interpolation. † denotes a statistically significant difference obtained over the baseline system (Sys. 1). }}
\scriptsize
\begin{center}
\vspace{-0.6cm}
\scalebox{0.9}[0.9]{
\begin{tabular}{c|c|p{4mm}p{4mm}p{4mm}p{4mm}|l}
\hline
\multirow{2}{*}{Sys.} & \multirow{2}{*}{Model Combination}   & \multicolumn{2}{c|}{Dev}                           & \multicolumn{2}{c|}{Eval}                          & \multirow{2}{*}{All} \\ \cline{3-6}
& \multicolumn{1}{c|}{}                             & \multicolumn{1}{c}{Inv} & \multicolumn{1}{c|}{Par} & \multicolumn{1}{c}{Inv} & \multicolumn{1}{c|}{Par} &                      \\ \hline \hline
- & CNN-TDNN    &  17.1     & 39.5     &  18.9     & 30.7      &28.3    \\
14 &  Conformer     &       16.0$\dag$  &    35.2$\dag$     &    15.3$\dag$  &     26.4$\dag$          &       25.3$\dag$  \\ 
16 &  Conformer   &      16.1$\dag$    &    36.1$\dag$     &    \textbf{14.7}$\dag$     &         27.6$\dag$    &   25.9$\dag$      \\ \hline
17 & CNN-TDNN $\rightarrow$ Sys.14   &      \textbf{15.0}$\dag$     &     \textbf{33.7}$\dag$    &    16.0   &  \textbf{25.5}$\dag$    &    \textbf{24.2}$\dag$   \\
18 & CNN-TDNN $\rightarrow$ Sys.16   &     15.4$\dag$  &    34.6$\dag$     &   15.7    &    25.9$\dag$  &   24.8$\dag$     \\
\hline
\end{tabular}
}
\vspace{-0.4cm} 
\end{center}
\end{table}
\vspace{-0.1cm}
\subsection{System Combination}
Fundamental modeling differences between hybrid and end-to-end ASR models create large diversity and complementarity among them. This has in recent years drawn increasing interest in developing suitable combination approaches to exploit such complementarity within the speech community\cite{watanabe2017hybrid, sainath2019two, evermann2000posterior, li2019integrating}. 
To this end, this paper considers a multi-pass rescoring based system combination of hybrid CNN-TDNN and Conformer based ASR systems. 
A hybrid CNN-TDNN system featuring similar designs of our previous LF-MMI TDNN system \cite{ye2021development} (domain and Bayesian LHUC speaker adaptation) and additional 2-dimensional convolutional structures as the first 6 layers \cite{xie2021bayesian} was used to produce initial N-best (N=100) outputs before being rescored using the manually or NAS auto-configured domain and LHUC speaker adapted Conformer systems (Sys. 14, 16 in Table 2) using a 2-way cross system sequence level log-likelihood score interpolation \cite{myc2021twopass}. 
As is shown in the bottom section of Table 3, the 2-pass combined systems (Sys. 17, 18) produced further WER reductions of 1.1\% absolute (0.9-1.7\% absolute for participants in the evaluation set).

\begin{table}[]
\caption{\label{tab:table-comb}\textnormal{ASR WER\% and AD detection performance in terms of accuracy\%, precision\%, recall\% and F1 score\% obtained using the ground truth transcripts (GT), the baseline or ASR outputs of Table 2 and 3, as well as published result for participants of the ADReSS evaluation set. }}
\footnotesize
\centering
\vspace{-0.2cm} 
\scalebox{0.8}[0.8]{
\begin{tabular}{c|c|cccc}
\hline 
Sys. &  Eval Par. WER    & Acc   & Pre. & Rec. & F1 \\ \hline \hline
GT  & - &  87.5     & 90.9      & 83.3 & 87.0    \\ \hline 
\cite{syed2021automated} & -   & 91.7     & -      & - & -    \\
\cite{yuan2020disfluencies} &  -    &  89.6     &   -    &  - & -   \\
\cite{ye2021development}    &  33.2  &    87.5     & 82.1      & 95.8 & 88.5  \\ \hline
1 &  39.1  & 79.2 & 76.9 & 83.3 & 80.0   \\ \cline{1-1}
14  & 26.4  & \textbf{91.7}  & \textbf{91.7}	&91.7	 &	91.7 \\ \cline{1-1}
16  &  27.6  &  89.6  &   85.1   &   \textbf{95.8} & 90.2                  \\ \cline{1-1}
17 & 25.5 &     87.5 &    82.1    &   \textbf{95.8} & 88.5     \\ \cline{1-1}
18 &  25.9  &    \textbf{91.7}   &   88.5   & \textbf{95.8} &      \textbf{92.0}  \\ \hline 
\hline
\end{tabular}
}
\vspace{-0.1cm}
\end{table}
\section{AD Detection Performance}
In this section, the textual features separately extracted from the DementiaBank Pitt evaluation set recognition outputs produced by the baseline and best performing Conformer systems (Sys. 1, 14, 16 in Table 2, Sys 17, 18 in Table 3) and the hybrid CNN-TDNN system were fed into a Support Vector Machine (SVM) based AD detection system. Embedding features were produced by feeding ASR transcriptions into a pre-trained BERT \cite{devlin2018bert} or Roberta \cite{liu2019roberta}. This detection system was maximum-margin trained on the ADReSS training set, a subset of the Cookie session transcripts of the Pitt corpus of 108 recordings \cite{luz2020alzheimer}. The SVM outputs based on either BERT or Roberta embedding features alone were further fused via majority voting to produce the final AD detection decision. More details of the detection system can be found in \cite{li2021comparative,ywang2021exploring}. As is shown in Table 4, the AD detection performance based on multiple ASR system outputs (Sys. 14, 16, 18) outperformed that using the ground truth speech transcripts. Compared with the previously published results on the same task using ground truth speech transcripts \cite{syed2021automated,yuan2020disfluencies}, the overall comparable AD detection accuracy of 91.7\%, precision of 95.8\% and F1 score of 92.0\% were obtained using the transcripts produced by the 2-pass combined systems (Sys. 18). 
\vspace{-0.1cm}
\section{Conclusion}
The development of a state-of-the-art Conformer ASR system constructed using the DementiaBank Pitt corpus was presented in this paper. A range of techniques including NAS, domain and speaker adaptation and system combination with hybrid TDNN models were employed to improve the recognition performance of elderly speech. An overall significant WER reduction of 13.6\% absolute (34.8\% relative) was obtained over the baseline recipe configured Conformer system on the Pitt evaluation set of 48 elderly speakers. The AD detection performance using the textual features extracted from our best ASR outputs is comparable to using ground truth speech transcripts and produced an accuracy of 91.7\%. Tighter integration between ASR and AD detection will be investigated in future research. 
\vspace{-0.1cm}
\section{Acknowledgment}
This research is supported by Hong Kong RGC GRF grant No. 14200021, 14200218, 14200220, TRS T45-407/19N and Innovation \& Technology Fund grant No. ITS/254/19.
\clearpage

\bibliographystyle{IEEEtran}
\bibliography{jinbib}

\end{document}